# Critical Fields and Anisotropy of $NdO_{0.82}F_{0.18}FeAs$ Single Crystals

Ying Jia, Peng Cheng, Lei Fang, Huiqian Luo, Huan Yang, Cong Ren, Lei Shan, Changzhi Gu and Hai-Hu Wen*
Institute of Physics, Chinese Academy of Sciences, Beijing 100190, China

The newly discovered iron-based superconductors have stimulated enormous interests in the field of superconductivity. Since the new superconductor is a layered system, the anisotropy is a parameter with the first priority to know. Meanwhile any relevant message about the critical fields (upper critical field and irreversibility line) are essentially important. By using flux method, we have successfully grown the single crystals $NdO_{0.82}F_{0.18}FeAs$ at ambient pressure. Resistive measurements reveal a surprising discovery that the anisotropy $\Gamma = (m_c/m_{ab})^{1/2}$ is below 5, which is much smaller than the theoretically calculated results. The data measured up to 400 K show a continuing curved feature which prevents a conjectured linear behavior for an unconventional metal. The upper critical fields determined based on the Werthamer-Helfand-Hohenberg formula are $H_{c2}^{ab}$(T=0 K) ≈ 304 T and $H_{c2}^{c}$(T=0 K) ≈ 62-70 T, indicating a very encouraging application of the new superconductors.

* To whom the correspondence should be addressed: hhwen@aphy.iphy.ac.cn

The newly discovery of superconductivity at 26 K in fluorine doped LaOFeAs generates enormous interests in the community of superconductivity[1]. The superconducting transition temperature has been raised to 55 K shortly[2-6] and superconductivity has been found both in electron doped and hole doped systems[7]. Meanwhile many experimental and theoretical results have been carried out leading to some important concluding remarks. For example, a spin-density-wave order at about 136 K was observed in undoped LaOFeAs system[8-10]; unconventional pairing symmetry or line nodes were discovered in the superconducting state by low temperature specific heat[11], point contact tunneling spectrum[12] and lower critical field[13], or NMR[14]; rather high upper critical fields have been observed[15-17], etc.. As far as we know, however, all these results were measured on polycrystalline samples. Some of them may suffer a change when data on single crystals are available. In this sense, data from single crystal samples are highly desired. Since the new system has also a layered structure, it is very interesting to have a precise evaluation on the critical fields, such as the upper critical field $H_{c2}(T)$ and the irreversibility line $H_{irr}(T)$ which measures the vortex melting transition, and how large the anisotropy is when compared with the cuprate superconductors. In this paper we report the successful fabrication of single crystals of $NdO_{0.82}F_{0.18}FeAs$ with $T_c$(onset) = 50 K and present the first hand fresh data on the new superconductors.

The crystals were made by flux method using NaCl as the flux. First the starting materials Nd grains (purity 99.95%) and As grains (purity 99.99%) were mixed in 1:1 ratio, ground and pressed into a pellet shape. Then it was sealed in an evacuated quartz tube and followed by burning at 800 °C for 10 hours. The resultant pellet was smashed and ground together with the $NdF_3$ powder (purity 99.95%), $Fe_2O_3$ powder (purity 99.9%) and Fe powder (purity 99.9%) in stoichiometry as the formula $NdO_{0.82}F_{0.18}FeAs$. Again it was pressed into a pellet and put together with NaCl powder of mass ratio (NaCl : NdOFFeAs = 10: 1) and sealed in an evacuated quartz tube and burned at about 1050 °C for 10 days. Then it was cooled down at a rate of 3

°C/hour to 850 °C and followed with a quick cooling down by shutting off the power of the furnace. The resulting product is a pullet with dark color. It contains mainly the plate-like small crystals. In the upper panel of Figure.1, we show the X-ray diffraction (XRD) patterns for the polycrystalline sample after growth mentioned above. One can see that all main peaks can be indexed by a tetragonal structure with a = b = 3.962 Å, c = 8.555 Å. The two small peaks marked with asterisks can be indexed to the structure of impurity phase NdClO. The bottom-right picture shows a result of scanning-electron-microscopy (SEM) from which one can see that our sample contains only plate-like crystals with lateral sizes ranging from 5 to 30 micrometers and thickness of about 1-5 micrometers. This picture shows that our sample is very different from others grown by solid reaction method, in which only grains with random and un-regular shapes were observed[(17)]. At some positions of the sample, we can get crystals as large as 50 micrometer. The left-bottom panel of Figure 1 shows the Energy Dispersive X-ray (EDX) microanalysis spectrum taken on one of the crystals. From the EDX analysis we found that this crystal contains only the right components Nd, Fe, As O and F, without any other elements. The composition of Nd:Fe:As on the crystals is found to be close to 1:1:1, but the exact composition of O and F cannot be resolved due to the light mass.

From the bulk sample, we can pick up the small crystals and make electric contacts on them using the Pt deposition technique of Focused-Ion-Beam (FIB). In the inset of Figure 2, one crystal with four Pt leads is shown. The surface of the crystal looks rather flat with a resolved view of layered structure at the edge. The main panel of Figure 2 shows the temperature dependence of resistivity measured up to 400 K. A dotted line in low temperature region is drawn to view the magnitude of the residual resistivity. One can see that the residual resistivity at T = 0 K is about $\rho_0 = 0.127$ mΩcm which is about 6.4 times smaller than that at 400 K. On another crystal with $\rho_0 = 0.028$ mΩcm and the ratio $\rho_{400K}/\rho_0 = 19$, the superconducting transition temperature is about $T_c$(onset) = 52 K, i.e., only 2K higher compared with the present sample. This may suggest that the impurity scattering is not very essential in controlling the superconducting transition temperatures. Another very interesting observation here is that the resistivity data exhibit a continuing curved feature up to 400 K without showing a hypothesized linear behavior expected by some models for unconventional metal, such as the case for optimally doped cuprates in high temperature region[18]. The same behavior occurs for other two crystals (not shown here) with similar $T_c$. Therefore we cannot rule out the possibility that the electron-phonon scattering dominates the normal state electric conduction and the Ioffe-Regel limit is satisfied beyond 400 K. This indicates that the scattering time $\tau$ is larger than $a/v_F = 3.07\times10^{-15}$s below 400 K when taking the Fermi velocity $v_F=1.3\times10^{7}$ cm/s[19] and the in-plane lattice constant a = 3.962 Å.

In Figure 3 we present an enlarged view of the temperature dependence of resistivity with the magnetic field applied alone ab-plane or c-axis. The resistive transition under zero-field shows a rather sharp transition. By applying a magnetic field along c-axis the transition becomes broadened gradually showing a strong vortex motion. There is a small "knee" on the resistive transition curve under a magnetic field. This kind of knee structure on the transition curve has actually been observed on cuprate[20] or $MgB_2$[21] single crystals and were ascribed to the possible first order melting of vortex lattice, or surface barrier for flux entry. When the field is applied along ab-plane, the resistive transition curve broadens slowly showing a strong vortex pinning in this configuration. This may suggest that the intrinsic pinning occurs also in the iron-based superconductors as in the cuprates. Details about the investigation on vortex motion will be presented separately.

From the resistive transitions shown in Figure 3, one can see that the curve near the onset transition is rounded, showing certain effect from the critical fluctuation. In order to determine the upper critical field, we draw a straight line based on the normal state data (shown in Fig.3 by the red solid lines) and adopt a criterion of $95\%\rho_n(T)$ instead of using the very onset point, such as $99\%\rho_n(T)$. The upper critical fields for H||ab and H||c-axis are determined in this way and shown in Figure 4. It is clear that the curve $H_{c2}^{ab}(T)$ is very steep with a slope of -d$H_{c2}^{ab}$/dT|$_{Tc}$ = 9T / K. However, when the magnetic field is applied along c-axis, the curve $H_{c2}^{c}(T)$ exhibits a much smaller slope and a positive curvature near $T_c$ with a slope of -d$H_{c2}^{c}$/dT|$_{Tc}$ = 1.85T / K. This positive curvature was also observed in the cuprate superconductors(22,23). However, the average slope for the configuration of H||c-axis is about 2.09 T / K. By using the Werthamer-Helfand- Hohenberg formula(24) $H_{c2}(0)=-0.69[dH_{c2}/dT]T_c$, and taking $T_c$ = 49 K, the values of upper critical fields are: $H_{c2}^{ab}(0)$ = 304 T, $H_{c2}^{c}(0)$ ranges from 62 T to 70 T depending on which slope at $T_c$ is taken. In Figure 4 the irreversibility lines (defined as 1%$\rho_n$) for field parallel or perpendicular to the ab-planes are also presented, one can see that the irreversibility field is rather high comparing to that in $MgB_2$. These high values for upper critical field indicate a very perspective application of this new superconductor.

Finally we have a discussion on the anisotropy of this new superconductor. According Lawrence-Doniach model(25), the anisotropy Γ and a simple relation between the mass matrix element and the upper critical field is given by

$$\Gamma = \left( m_c / m_{ab} \right)^{1/2} = \xi_{ab} / \xi_c = H_{c2}^{ab} / H_{c2}^{c} \qquad (1)$$

Here $m_c$ and $m_{ab}$ are the mass matrix elements when the electrons move along c-axis and ab-plane respectively, $\xi_c$ and $\xi_{ab}$ are the coherence length along c-axis and ab-plane. Using $H_{c2}^{ab}$ = 304 T, $H_{c2}^{c}$=62-70 T, we get an anisotropy of $\Gamma \approx$ 4.9-4.34. For a much cleaner sample [$\rho_0 = 0.028$ mΩcm, $\rho_{400K} / \rho_0 = 19$ and $T_c$(onset) = 52 K] we found that $\Gamma \leq 6$. This value of anisotropy is even below $\Gamma \approx 7$ for $YBa_2Cu_3O_7$ single crystals(26). This is again a very important result which indicates that the anisotropy is not that strong and it is very optimistic for the application. Based on this analysis we get the mass ratio $m_c/m_{ab}$= 19 – 24. According to the band structure calculations by Lebègue[27] and by Singh and Du[19], the Fermi surface consists of five sheets: two electron cylinders centered around M-A line with high velocity, two hole cylinders around Γ-Z line with lower velocity, and an additional heavy 3D hole pocket which is centered at Z. It is supposed that the electron cylinders have higher velocity and make the dominant contribution to the in-plane electrical conductivity. Taking these two electron pockets into account, the calculated Fermi velocities are $v_{xx} = v_{yy} = 1.30\times10^7$ cm/s and $v_{zz} = 0.34\times10^7$ cm/s which yield a ratio $v_{xx} / v_{zz} = 15$. Because $v_F^2 = 2\varepsilon_F / \hbar^2 m$ with $\varepsilon_F$ the Fermi energy, one expects that $m_c/m_{ab}$ = 225! This is certainly much larger than what we get from the experiment $m_c/m_{ab}$= 19 – 24. The situation becomes even worse with more doped electrons, since according to the theory[19] the anisotropy is increased when the system is doped away from the parent phase. The large discrepancy between the theoretical calculation and the experiment data may suggest that the 3D hole pocket and the 2D hole cylinders all collectively contribute to the electron conduction and superconductivity[28].

29. We are grateful for fruitful discussions with Qian Niu. YJ, PC and LF contribute equally to this work. HHW designed and coordinated the whole work, analyzed the data and wrote the paper. This work is supported by the Natural Science Foundation of China, the Ministry of Science and Technology of China (973 project No: 2006CB60100, 2006CB921107, 2006CB921802), and Chinese Academy of Sciences (Project ITSNEM).

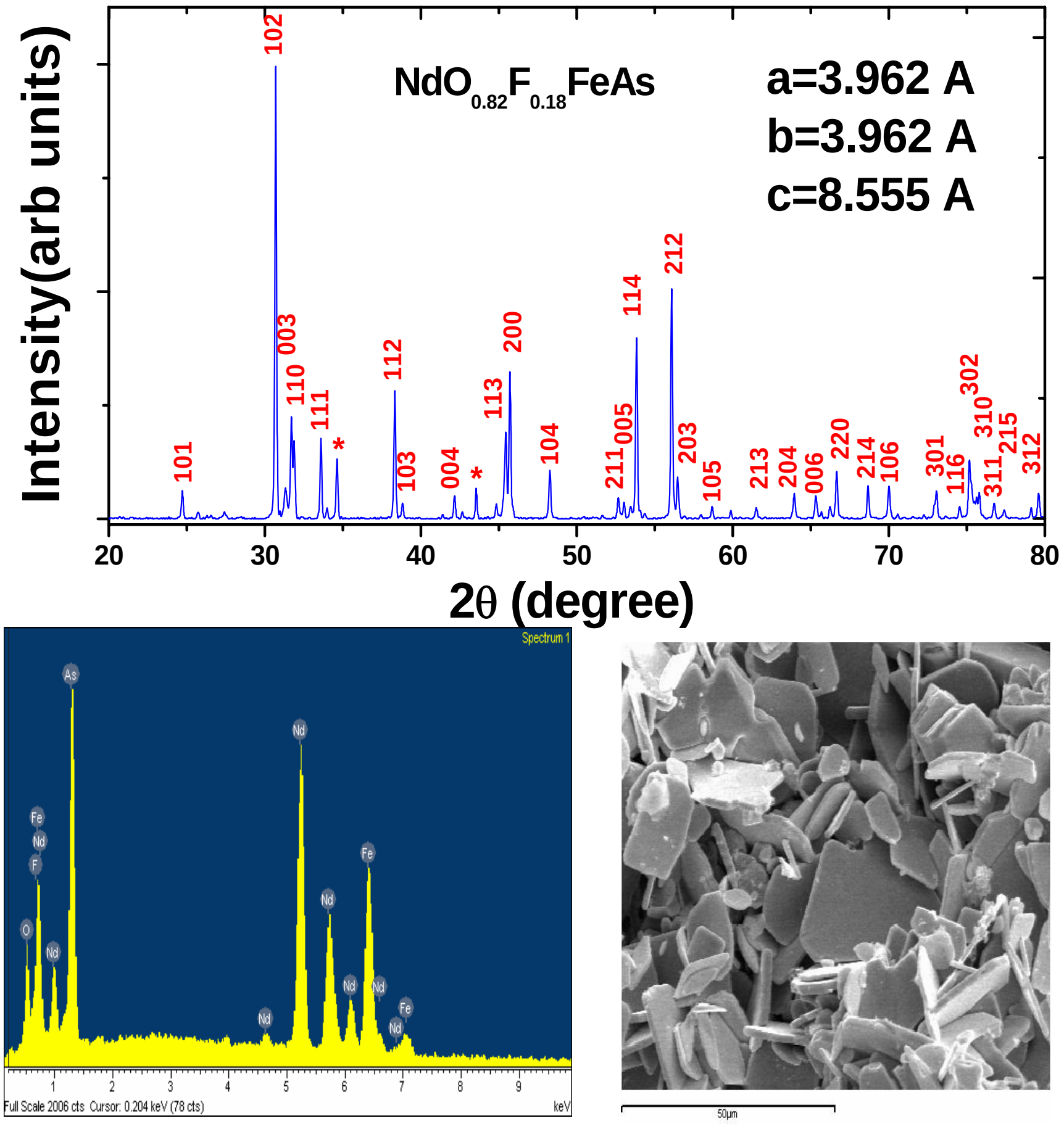

Fig.1 Top view: X-Ray diffraction pattern of the polycrystalline sample which contains many plate-like crystals in lateral sizes of 5-50 micrometer as shown in the right-bottom picture of scanning electron microscopy. The two tiny peaks marked with asterisks are indexed to NdClO impurity phase. Picture in left-bottom shows the Energy Dispersive X-ray microanalysis spectrum taken on one crystal.

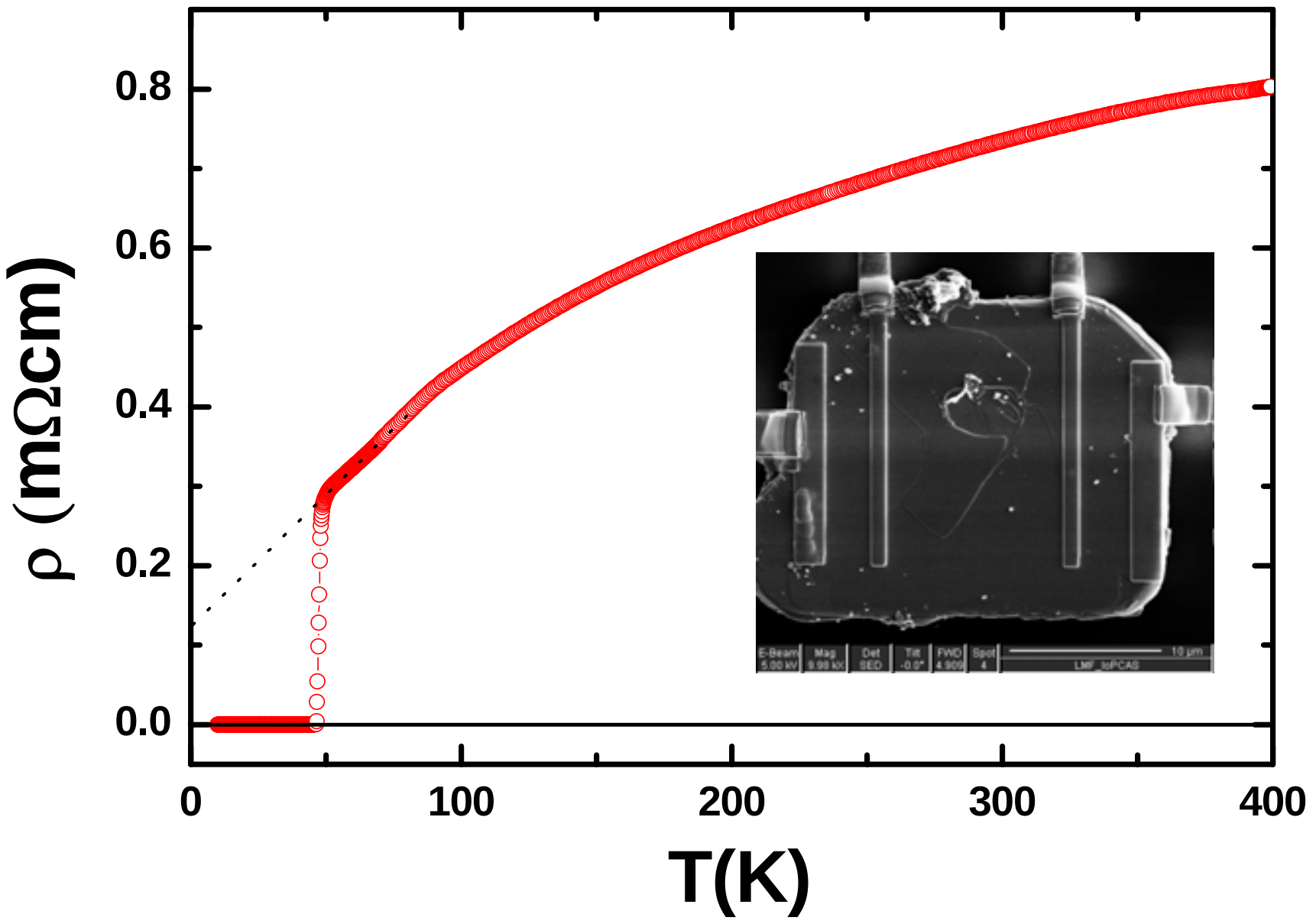

Fig.2 The temperature dependence of resistivity measured on the $NdO_{0.82}F_{0.18}FeAs$ crystal shown in the inset. The electric contacts of Pt metal were made by using the Pt deposition technique of Focused-Ion-Beam. The dotted line is a guide to the eye showing a residual resistivity of 0.127 mΩcm.

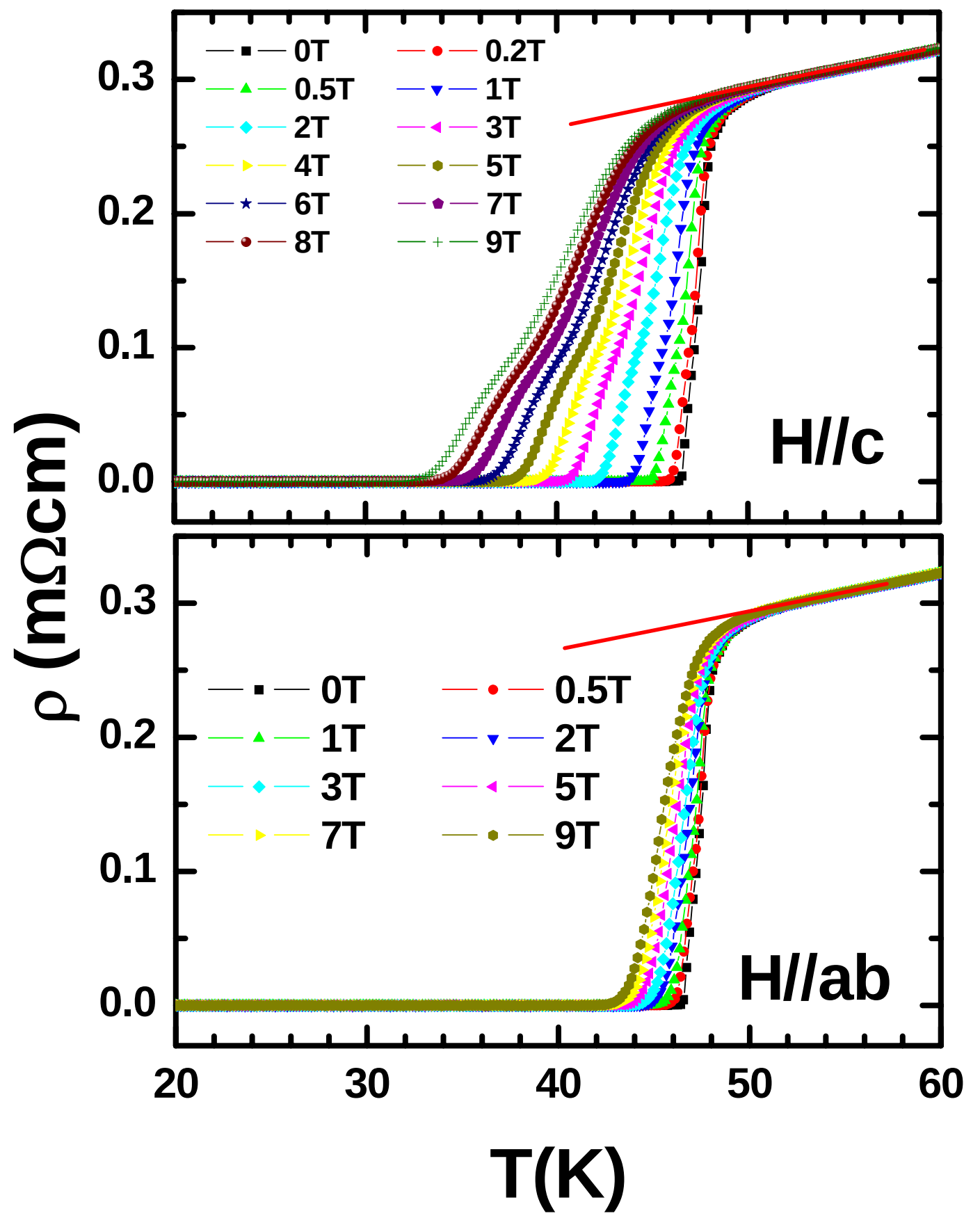

Fig.3 Temperature dependence of resistivity under magnetic fields up to 9 T when (a) H||c and (b) H||ab. One can see that the resistive broadening is more pronounced when the magnetic field is parallel to c-axis, showing a stronger vortex motion in this configuration. The small "knee" in the curve may be attributed to the first order melting of vortex lattice, or the surface effect of the crystal for flux entry, both can be observed quite often on superconducting single crystals (see text).

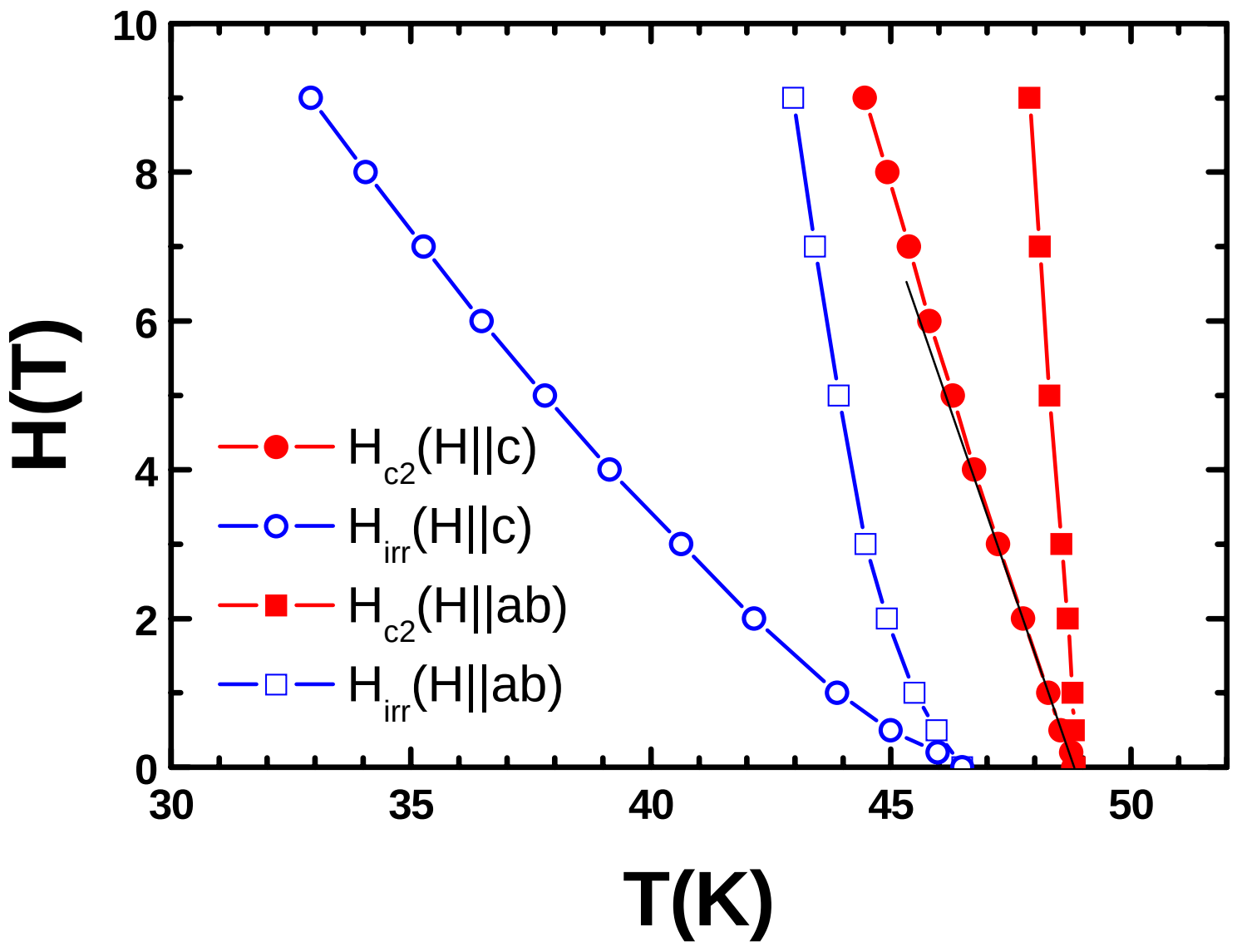

Fig.4 The mixed state phase diagram of the $NdO_{0.82}F_{0.18}FeAs$ crystal. The red filled symbols show the upper critical fields determined from the onset transition point (95 % $\rho_n$, see text). The blue open symbols represent the irreversibility line $H_{irr}(T)$. The dark linear line on the curve $H_{c2}^{c}(T)$ indicates the lowest slope near $T_c$.